\documentclass[preprintnumbers,amsmath,amssymbm,prd]{revtex4}
\usepackage{epsfig}
\usepackage{graphicx}

\begin{document}
\title{The Hawking evaporation process of rapidly-rotating black holes: An almost continuous cascade of gravitons}
\author{Shahar Hod}
\affiliation{The Ruppin Academic Center, Emeq Hefer 40250, Israel}
\affiliation{ } \affiliation{The Hadassah Institute, Jerusalem
91010, Israel}
\date{\today}

\begin{abstract}
\ \ \ It is shown that rapidly-rotating Kerr black holes are
characterized by the dimensionless ratio
$\tau_{\text{gap}}/\tau_{\text{emission}}=O(1)$, where
$\tau_{\text{gap}}$ is the average time gap between the emission of
successive Hawking quanta and $\tau_{\text{emission}}$ is the
characteristic timescale required for an individual Hawking quantum
to be emitted from the black hole. This relation implies that the
Hawking cascade from rapidly-rotating black holes has an almost
continuous character. Our results correct some inaccurate claims
that recently appeared in the literature regarding the nature of the
Hawking black-hole evaporation process.
\end{abstract}
\bigskip
\maketitle

\section{Introduction}

Working within the framework of semi-classical general relativity,
Hawking \cite{Haw} has revealed that black holes are actually not
completely black. In particular, it was shown \cite{Haw} that black
holes are characterized by thermally distributed emission spectra
\cite{Notegrey}. This intriguing finding is certainly one of the
most important theoretical predictions of modern physics.

It has recently been pointed out \cite{Vis} that the Hawking
radiation flux out of spherically-symmetric (non-rotating)
Schwarzschild black holes is extremely sparse (see
\cite{BekMuk,Mak,Hod1} for earlier discussions of this black-hole
property). In particular, denoting by $\tau_{\text{gap}}$ the
average time gap between the emissions of successive black-hole
quanta, and by $\tau_{\text{emission}}$ the characteristic timescale
required for an individual Hawking quantum to be emitted from the
black hole, one finds the characteristic dimensionless ratio
\cite{Vis,BekMuk,Mak,Hod1}
\begin{equation}\label{Eq1}
\eta\equiv{{\tau_{\text{gap}}}\over{\tau_{\text{emission}}}}=O(10^2)\
\end{equation}
for Schwarzschild black holes.

The relation (\ref{Eq1}) implies that the quantum radiation flux of
Schwarzschild black holes is indeed sparse. Namely, the average time
gap between the emissions of successive Hawking quanta out of a
Schwarzschild black hole is very large on the timescale
$2\pi/\omega$ [see Eq. (\ref{Eq8}) below] set by the characteristic
energy (frequency) of these emitted quanta. Thus, one may safely
conclude that the Hawking quanta of an evaporating Schwarzschild
black hole are typically emitted from the black hole one at a time
\cite{Vis,BekMuk,Mak,Hod1}.

It was also claimed in \cite{Vis} that adding angular momentum to
the black hole makes the dimensionless ratio $\eta$ even {\it
larger}, thus making the Hawking emission spectra of rotating Kerr
black holes even more sparse than the corresponding emission
spectrum (\ref{Eq1}) of the non-rotating Schwarzschild black hole.
Namely, it was claimed in \cite{Vis} that
\begin{equation}\label{Eq2}
\eta_{\text{Kerr}}\geq \eta_{\text{Schwarzschild}}\  .
\end{equation}

As we shall show in this paper, the claim (\ref{Eq2}) of \cite{Vis}
is actually erroneous. In particular, explicit calculations to be
carried below reveal that, for rapidly-rotating Kerr black holes,
$\eta(\bar a)$ is a {\it decreasing} function of the dimensionless
black-hole angular momentum $\bar a$ \cite{Noteaa}. Moreover, as we
shall show below, near-extremal Kerr black holes are actually
characterized by the relation $\eta(\bar a\to 1)=O(1)$.

\section{The Hawking evaporation process of rapidly-rotating Kerr black
holes}

We study the Hawking emission of gravitational quanta by rotating
Kerr black holes. The Bekenstein-Hawking temperature of a Kerr black
hole and the angular velocity of its horizon are respectively given
by the relations \cite{Noteunit}
\begin{equation}\label{Eq3}
T_{\text{BH}}={{\hbar(r_+-r_-)}\over{4\pi(r^2_++a^2)}}\ \ \ \
\text{and}\ \ \ \ \Omega_{\text{H}}={{a}\over{2Mr_+}}\  ,
\end{equation}
where $r_{\pm}=M\pm (M^2-a^2)^{1/2}$ are the (inner and outer)
horizon radii of the black hole.

The Hawking emission rate (that is, the number of quanta emitted per
unit of time) out of a rotating Kerr black hole is given by the
Hawking relation \cite{Haw,Page}
\begin{equation}\label{Eq4}
{\cal \dot
N}\equiv{{dN}\over{dt}}={{\hbar}\over{2\pi}}\sum_{l,m}\int_0^{\infty}
d\omega{{\Gamma}\over{e^x-1}}\  ,
\end{equation}
where $x\equiv\hbar(\omega-m\Omega_{\text{H}})/T_{\text{BH}}$. Here
$l$ and $m$ (with $l\geq |m|$) are respectively the spheroidal
harmonic index and the azimuthal harmonic index of the emitted
quanta, and $\Gamma=\Gamma_{lm}(\omega;\bar a)$ are the
frequency-dependent gray-body factors \cite{Page}. These
dimensionless absorption probabilities quantify the imprint of
passage of the emitted black-hole quanta through the effective
curvature potential which characterizes the black-hole spacetime.

The Hawking emission rate ${\cal \dot N}(\bar a)$ [see Eq.
(\ref{Eq4})] can be computed along the lines of the numerical
procedure described in \cite{Page}. In particular, one finds that,
for rapidly-rotating Kerr black holes, the Hawking emission spectrum
is greatly dominated by gravitational quanta with the angular
indices \cite{Page,Notegr}
\begin{equation}\label{Eq5}
l=m=s=2\  .
\end{equation}

Moreover, the characteristic thermal factor that appears in the
denominator of (\ref{Eq4}) implies that, for rapidly-rotating
(near-extremal, $T_{\text{BH}}\to 0$) black holes, the emission of
high energy quanta with $\omega>m\Omega_{\text{H}}$ is exponentially
suppressed. Thus, for rapidly-rotating Kerr black holes, the Hawking
emission spectra are effectively restricted to the regime
\begin{equation}\label{Eq6}
0\leq\omega\lesssim m\Omega_{\text{H}}+O(T_{\text{BH}}/M^2)\  .
\end{equation}

The reciprocal of the black-hole emission rate,
\begin{equation}\label{Eq7}
\tau_{\text{gap}}={{1}\over{{\cal \dot N}}}\  ,
\end{equation}
determines the average time gap between the emissions of successive
Hawking quanta. On the other hand, the characteristic timescale
required for each individual Hawking quantum to be emitted from the
black hole, $\tau_{\text{emission}}$, can be bounded from below by
the time-period it takes to the corresponding emitted wave field to
complete a full oscillation cycle. That is \cite{Notew1},
\begin{equation}\label{Eq8}
\tau_{\text{emission}}\geq\tau_{\text{oscillation}}={{2\pi}\over{\bar\omega}}\
,
\end{equation}
where $\bar\omega$ is the characteristic (average) frequency of the
emitted Hawking quanta \cite{Notew2}.

In Table \ref{Table1} we display the characteristic dimensionless
ratio ${{\tau_{\text{gap}}}/{\tau_{\text{oscillation}}}}$ for
rapidly-rotating Kerr black holes \cite{Notebo}. One finds that
$\eta(\bar a)$ is a {\it decreasing} function of the dimensionless
black-hole angular momentum $\bar a$. In particular, we find that
near-extremal Kerr black holes are characterized by the relation
\begin{equation}\label{Eq9}
\eta(\bar a\to 1)=O(1)\ .
\end{equation}

\begin{table}[htbp]
\centering
\begin{tabular}{|c|c|c|c|c|c|}
\hline $\bar a\equiv J/M^2$ & \ \ 0.90 \ \ & \ \ 0.96 \ \ & \ \ 0.99
\ \ & \ \ 0.999 \ \ & \ \ 1.0\ \ \ \\
\hline \ \ ${{\tau_{\text{gap}}}/{\tau_{\text{oscillation}}}}$
\ \ &\ 13.5\ \ &\ \ 5.5\ \ &\ 2.5\ \ &\ 1.5\ \ &\ \ 1.2\ \ \ \\
\hline
\end{tabular}
\caption{The characteristic dimensionless ratio
${{\tau_{\text{gap}}}/{\tau_{\text{oscillation}}}}$ of
rapidly-rotating Kerr black holes. Here $\tau_{\text{gap}}$ is the
average time gap between the emission of successive Hawking quanta
[see Eq. (\ref{Eq7})] and $\tau_{\text{oscillation}}$ is the
characteristic oscillation period of the emitted wave field [see Eq.
(\ref{Eq8})]. One finds that $\eta(\bar a)$ is a {\it decreasing}
function of the dimensionless black-hole angular momentum $\bar a$.
In particular, near-extremal Kerr black holes are characterized by
the relation $\eta(\bar a\to 1)=O(1)$.} \label{Table1}
\end{table}

\section{Summary}

It has long been known \cite{Vis,BekMuk,Mak,Hod1} that the Hawking
radiation flux out of Schwarzschild black holes is extremely sparse.
In particular, these spherically-symmetric black holes are
characterized by the large dimensionless ratio
$\eta\equiv\tau_{\text{gap}}/\tau_{\text{emission}}=O(10^2)$ [see
Eq. (\ref{Eq1})]. As recently pointed out in \cite{Vis}, this
relation implies that the individual Hawking quanta emitted from a
Schwarzschild black hole are well separated in time.

It was recently claimed in \cite{Vis} that adding angular momentum
to the emitting black hole makes the dimensionless ratio $\eta$ even
larger [see Eq. (\ref{Eq2})], thus making the Hawking radiation
spectra of rotating Kerr black holes even sparser than the
corresponding emission spectrum of the (non-rotating) Schwarzschild
black hole.

In this brief report we have explicitly shown that the claim
(\ref{Eq2}) made in \cite{Vis} is actually erroneous. In particular,
explicit calculations reveal that, for rapidly-rotating Kerr black
holes, $\eta(\bar a)$ is actually a {\it decreasing} function of the
dimensionless black-hole angular momentum $\bar a$. Moreover, we
have shown that near-extremal Kerr black holes are characterized by
the relation $\eta(\bar a\to 1)=O(1)$.

The relation $\tau_{\text{gap}}/\tau_{\text{emission}}=O(1)$ [see
Eq. (\ref{Eq9})] implies that the Hawking cascade of gravitons from
rapidly rotating Kerr black holes has an almost {\it continuous}
character. Stated in a more picturesque way, we can say that, on
average, there is a gravitational quantum leaving the
(rapidly-rotating) black hole at any given moment of time.


\bigskip
\noindent {\bf ACKNOWLEDGMENTS}
\bigskip

This research is supported by the Carmel Science Foundation. I thank
Yael Oren, Arbel M. Ongo, Ayelet B. Lata, and Alona B. Tea for
stimulating discussions.


\begin{thebibliography}{99}
\bibitem{Notegrey} It is important to emphasize that the black-hole thermal spectrum
is distorted by the curvature potential which surrounds the black
hole, see Eq. (\ref{Eq4}) below.
\bibitem{Noteaa} Here $\bar a\equiv J/M^2$ is the dimensionless angular momentum of the
Kerr black hole.
\bibitem{Noteunit} We use gravitational units in which $G=c=k_{\text{B}}=1$.
\bibitem{Notegr} The last equality in (\ref{Eq5}) refers to gravitational quanta,
which are characterized by the spin parameter $s=2$.
\bibitem{Notew1} Note that there is a factor $2\pi$ difference
between our definition of $\tau_{\text{oscillation}}$ and the
corresponding definition used in \cite{Vis}. We believe that the
time period ${{2\pi}/{\omega}}$ required by the emitted wave field
to complete a full oscillation cycle [see Eq. (\ref{Eq8})] provides
a natural lower bound on the characteristic emission timescale
$\tau_{\text{emission}}$ of the corresponding Hawking quantum.
\bibitem{Notew2} Using in (\ref{Eq8}) the peak frequency $\omega_{\text{peak}}$ of the Hawking emission spectrum
instead of the average frequency $\bar\omega$ of the Hawking
emission spectrum would merely change the dimensionless ratio $\eta$
by a factor of order unity.
\bibitem{Notebo} It is worth emphasizing again that this ratio
provides an upper bound on the actual value of $\eta(\bar
a)\equiv{{\tau_{\text{gap}}}/{\tau_{\text{emission}}}}$, see Eq.
(\ref{Eq8}).
\end{thebibliography}
\end{document}